\documentclass[preprint,showpacs,preprintnumbers]{revtex4}
\usepackage{amsmath}
\usepackage{graphicx}
\usepackage{dcolumn}
\usepackage{bm}
\usepackage{epsfig}
\usepackage[T1]{fontenc}
\usepackage{ae,aecompl}

\begin{document}

\title{Multiple-relaxation-time lattice Boltzmann model for compressible
fluids}
\author{Feng Chen$^1$, Aiguo Xu$^2$\footnote{
Corresponding author. E-mail: Xu\_Aiguo@iapcm.ac.cn}, Guangcai Zhang$^2$,
Yingjun Li$^1$}
\affiliation{1, China University of Mining and Technology (Beijing), Beijing 100083 \\
2, National Key Laboratory of Computational Physics, \\
Institute of Applied Physics and Computational Mathematics, P. O. Box
8009-26, Beijing 100088, P.R.China }
\date{\today }

\begin{abstract}
We present an energy-conserving multiple-relaxation-time finite
difference lattice Boltzmann model for compressible flows. This
model is based on a 16-discrete-velocity model. The collision step
is first calculated in the moment space and then mapped back to the
velocity space. The moment space and corresponding transformation
matrix are constructed according to the group representation theory.
Equilibria of the nonconserved moments are chosen according to the
need of recovering compressible Navier-Stokes equations through the
Chapman-Enskog expansion. Numerical experiments showed that
compressible flows with strong shocks can be well simulated by the
present model. The used benchmark tests include (i) shock tubes,
such as the Sod, Lax, Colella explosion wave, collision of two
strong shocks and a new shock tube with high mach number, (ii)
regular and Mach shock reflections, (iii) shock wave reaction on
cylindrical bubble problems, and (iv)Couette flow. The new model
works for both low and high speeds compressible flows. It contains
more physical information and has better numerical stability and
accuracy than its single-relaxation-time version.
\end{abstract}

\pacs{47.11.-j, 51.10.+y, 05.20.Dd \\
\textbf{Keywords:}lattice Boltzmann method; compressible flows;
multiple-relaxation-time; von Neumann stability analysis} \maketitle

\section{Introduction}

The Lattice Boltzmann (LB) method is an innovative numerical scheme
originated from Lattice Gas Automata (LGA)\cite{1} and aim to simulate
various hydrodynamics\cite{2}. The LB method was introduced to overcome some
serious deficiencies of LGA, such as intrinsic noise, limited values of
transport coefficients, non-Galilean invariance, and implementation
difficulty in three dimensions. In the past two decades, most of the LB
models are based on the famous Bhatnagar-Gross-Krook (BGK) approximation\cite%
{5} where a Single Relaxation Time (SRT) is used. Due to its validity and
simplicity, the SRT LB method has been widely used to simulate various fluid
flow problems, such as the multiphase flow\cite%
{ShanChen,Swift,XGL1,XGL2,XGL6,XSB,SS2007,Xu2009}, magnetohydrodynamics\cite%
{PRL1991,7,8,PRA1991,PRE2002,PRE2004,CiCP2008}, flows through porous media%
\cite{9,10,10a} and thermal fluid dynamics\cite{11,11a,12}, etc.

However, the extreme simplicity of the SRT leads also to some constraints
for the SRT LB model. For example, the simulation will be unstable when the
relaxation time $\tau $ is close to 0.5, the model works only for low Mach
number flows. One possible remedy is to use the Multiple Relaxation Time
(MRT) method\cite{HSB,13}. In real fluid the equilibrating rates of mass,
momentum, energy, etc. are generally different. This difference can be
manifested by the non-unique adjustable parameters in the MRT LB model. In
contrast to the SRT model, the MRT version has much more adjustable
parameters and degrees of freedom. The relaxation rates of various processes
owing to particle collisions may be adjusted independently. The main
strategy of the MRT LB scheme is that the collision step is first calculated
in the moment space and then mapped back to the velocity space. The
advection step is still computed in the velocity space. In many cases, it
has been shown by Luo, et al\cite{17,19} that the MRT LB model has better
numerical stability.

Recently, the MRT LB method has attracted considerable interest and much
progress has been achieved. For example, MRT models for viscoelastic fluids%
\cite{15,16,16a}, multiphase flows\cite{20,200}, flow with free surfaces\cite%
{18}, etc. were developed; optimal boundary condition for MRT LB was composed%
\cite{14}. To simulate system with temperature field, Luo, et al.\cite{19}
suggested a hybrid thermal MRT LB model. These models work only for nearly
incompressible fluids with very low Mach number.

LB community has long been attempting to construct models for compressible
fluids\cite{Alexander1,li1,sun1,yan1}. Alexander and Chen et al.\cite%
{Alexander1} constructed a model where the sound speed is adjustable so that
the Mach number can be enhanced. Li, et al.\cite{li1} gave a model by
reforming the velocity space. Sun, et al.\cite{sun1} formulated adaptive LB
models where the particle velocities are determined by the mean velocity and
internal energy. Yan, et al. \cite{yan1} proposed
three-speed-three-energy-level models. Besides the standard LB mentioned
above, some researchers have also tried to develop Finite Difference (FD) LB
for compressible fluids\cite{20a,21a,2a}, but in the real simulations the
accessible Mach number is still not large. The model introduced by Kataoka
and Tsutahara\cite{21a} uses only sixteen discrete velocities and hence has
a high computational efficiency.

The low-Mach number constraint is generally related to a numerical
stability problem. The latter has been partly addressed by a number
of techniques, such as the entropic method\cite{16+,17+}, the fix-up
scheme\cite{16+,18+}, Flux-limiters\cite{Sofonea1} and
dissipation\cite{43,Brownlee1} techniques. In existing SRT models,
it seems that the most effective solution to overcome the low Mach
number constraint is to introduce artificial viscosity. But with the
artificial viscosity, some fundamental kinetics are not very clear.
In many cases, the MRT formulation has been shown to offer improved
numerical stability, and provide additional physics. In this paper
we present an energy-conserving multiple relaxation time finite
difference lattice Boltzmann model for compressible flows with high
Mach number. This model is based on the one proposed by Kataoka and
Tsutahara\cite{21a}. The moment space and transformation matrix are
constructed according to the group representation theory. Equilibria
of the nonconserved moments in the moment space are chosen when
recovering compressible Navier-Stokes (NS) equations through the
Chapman-Enskog (CE) expansion.

This paper is organized as follows. In Sect. II a brief review to the MRT LB
model is presented. In Sect. III the new model is constructed. The von
Neumann stability analysis is given in Sect. IV. Section V shows the
numerical tests and some simulation results. Section VI provides a summary
and concludes the paper.


\section{ Brief review of the MRT LB model}

The evolution of the distribution function $f_{i}$\ for the particle
velocity $v_{i}$\ is governed by the following equation:
\begin{equation}
\frac{\partial f_{i}}{\partial t}+v_{i\alpha }\frac{\partial f_{i}}{\partial
x_{\alpha }}=-\mathbf{S}_{ik}\left[ f_{k}-f_{k}^{eq}\right] \text{,}
\label{1}
\end{equation}%
where $f_{i}$ ($f_{i}^{eq}$)\ is the particle (equilibrium) distribution
function, $v_{i}$ represents a group of particle velocities, subscript $i$
indicates the particle's direction, $i=1,\ldots ,N$, $N$\ is the number of
discrete velocities, the subscript $\alpha $\ indicates $x$\ or $y$
component, $\mathbf{S}$ is the collision matrix. The equation reduces to the
usual lattice BGK equation if all the relaxation parameters are set to be a
single relaxation time $\tau $, namely $\mathbf{S}=\frac{1}{\tau }\mathbf{I}$%
, where $\mathbf{I}$ is the identity matrix.

The discrete (equilibrium) distribution function $f_{i}$ ($f_{i}^{eq}$) in
Eq. \eqref{1} can be listed with the following matrixes:
\begin{subequations}
\begin{equation}
\mathbf{f}=\left( f_{1},f_{2},\cdots ,f_{N}\right) ^{T}\text{,}  \label{2a}
\end{equation}%
\begin{equation}
\mathbf{f}^{eq}=\left( f_{1}^{eq},f_{2}^{eq},\cdots ,f_{N}^{eq}\right) ^{T}%
\text{,}  \label{2b}
\end{equation}%
where $T$ is the transpose operator.

Given a set of discrete velocities $v_{i}$, and corresponding distribution
functions $f_{i}$, we can get a velocity space $S^{V}$, spanned by discrete
velocities $v_{i}$, and a moment space $S^{M}$, spanned by moments of
particle distribution function $f_{i}$, where $i=1,\cdots ,N$. Similarly, we
also express the moments of distribution function with the column matrix: $%
\hat{\mathbf{f}}=\left( \hat{f}_{1},\hat{f}_{2},\cdots ,\hat{f}_{N}\right)
^{T}$, where $\hat{f}_{i}=m_{ij}f_{j}$, $m_{ij}$ is an element of the matrix
$\mathbf{M}$ and is a polynomial of discrete velocities. Obviously, the
moments are simply linear combination of distribution functions$\ f_{i}$,
and the mapping between moment space and velocity space is defined by the
linear transformation $\mathbf{M}$, i.e., $\hat{\mathbf{f}}=\mathbf{Mf}$, $%
\mathbf{f=M}^{-1}\hat{\mathbf{f}}$, where $\mathbf{M}=\left(
m_{1},m_{2},\cdots ,m_{N}\right) ^{T},m_{i}=(m_{i1},m_{i2},\cdots ,m_{iN})$.

The LB simulation consists of two steps: the collision step and the
advection one. In the MRT LB method, the advection step is computed in the
velocity space. The collision step is first calculated in the moment space
and then mapped to the velocity space. So, the MRT LB equation can be
described as:
\end{subequations}
\begin{equation}
\frac{\partial f_{i}}{\partial t}+v_{i\alpha }\frac{\partial f_{i}}{\partial
x_{\alpha }}=-\mathbf{M}_{il}^{-1}\hat{\mathbf{S}}_{lk}(\hat{f}_{k}-\hat{f}%
_{k}^{eq})\text{,}  \label{3}
\end{equation}%
where $\hat{\mathbf{S}}=\mathbf{MSM}^{-1}=diag(s_{1},s_{2}, \cdots ,s_{N})$
is the diagonal relaxation matrix. $\hat{f}_{i}^{eq}$\ is the equilibrium
value of the moment $\hat{f}_{i}$. The moments can be divided into two
groups. The first group consists of the moments locally conserved in the
collision process, i.e. $\hat{f}_{i}=\hat{f}_{i}^{eq}$. The second group
consists of the moments not conserved, i.e. $\hat{f}_{i}\neq \hat{f}%
_{i}^{eq} $. The equilibrium $\hat{f}_{i}^{eq}$\ is a function of conserved
moments.


\section{Energy-conserving MRT LB model}

We use the two-dimensional discrete velocity model by Kataoka and Tsutahara%
\cite{21a} (see Fig. 1). It can be expressed as:
\begin{equation}
\left( v_{i1},v_{i2}\right) =\left\{
\begin{array}{cc}
\mathbf{cyc}:\left( \pm 1,0\right) \text{,} & \text{for }1\leq i\leq 4\text{,%
} \\
\mathbf{cyc}:\left( \pm 6,0\right) \text{,} & \text{for }5\leq i\leq 8\text{,%
} \\
\sqrt{2}\left( \pm 1,\pm 1\right) \text{,} & \text{for }9\leq i\leq 12\text{,%
} \\
\frac{3}{\sqrt{2}}\left( \pm 1,\pm 1\right) \text{,} & \text{for }13\leq
i\leq 16\text{,}%
\end{array}%
\right.  \label{4}
\end{equation}%
where \textbf{cyc} indicates the cyclic permutation.
\begin{figure}[tbp]
\center\includegraphics*[width=0.40\textwidth]{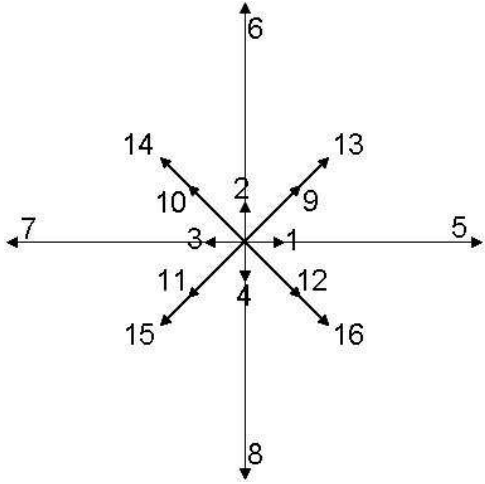}
\caption{Distribution of $\mathbf{v}_{i\protect\alpha }$ for the discrete
velocity model.}
\end{figure}

\subsection{Construction of transformation matrix $\mathbf{M}$}

The transformation matrix $\mathbf{M}$ is constructed according to the
irreducible representations of SO(2) group:
\begin{eqnarray*}
&&1\text{,} \\
&&\cos \theta \text{,}\sin \theta \text{,} \\
&&\sin ^{2}\theta +\cos ^{2}\theta \text{,}\cos 2\theta \text{,}\sin 2\theta
\text{,} \\
&&\cos \theta (\sin ^{2}\theta +\cos ^{2}\theta )\text{,}\sin \theta (\sin
^{2}\theta +\cos ^{2}\theta )\text{,}\cos 3\theta \text{,}\sin 3\theta \text{%
,} \\
&&(\sin ^{2}\theta +\cos ^{2}\theta )^{2}\text{,}\cos 2\theta (\sin
^{2}\theta +\cos ^{2}\theta )\text{,}\sin 2\theta (\sin ^{2}\theta +\cos
^{2}\theta )\text{,}\cos 4\theta \text{,}\sin 4\theta \text{,} \\
&&\cdots
\end{eqnarray*}

Let $v_{ix}$ and $v_{iy}$ play the roles of $\cos \theta $ and $\sin \theta $%
, respectively. Then we define
\begin{subequations}
\begin{equation}
m_{1i}=1\text{,}  \label{5a}
\end{equation}%
\begin{equation}
m_{2i}=v_{ix}\text{,}  \label{5b}
\end{equation}%
\begin{equation}
m_{3i}=v_{iy}\text{,}  \label{5c}
\end{equation}%
\begin{equation}
m_{4i}=(v_{ix}^{2}+v_{iy}^{2})/2\text{,}  \label{5d}
\end{equation}%
\begin{equation}
m_{5i}=v_{ix}^{2}-v_{iy}^{2}\text{,}  \label{5e}
\end{equation}%
\begin{equation}
m_{6i}=v_{ix}v_{iy}\text{,}  \label{5f}
\end{equation}%
\begin{equation}
m_{7i}=v_{ix}(v_{ix}^{2}+v_{iy}^{2})/2\text{,}  \label{5g}
\end{equation}%
\begin{equation}
m_{8i}=v_{iy}(v_{ix}^{2}+v_{iy}^{2})/2\text{,}  \label{5h}
\end{equation}%
\begin{equation}
m_{9i}=v_{ix}(v_{ix}^{2}-3v_{iy}^{2})\text{,}  \label{5i}
\end{equation}%
\begin{equation}
m_{10i}=v_{iy}(3v_{ix}^{2}-v_{iy}^{2})\text{,}  \label{5j}
\end{equation}%
\begin{equation}
m_{11i}=(v_{ix}^{2}+v_{iy}^{2})^{2}/4\text{,}  \label{5k}
\end{equation}%
\begin{equation}
m_{12i}=v_{ix}^{4}-6v_{ix}^{2}v_{iy}^{2}+v_{iy}^{4}\text{,}  \label{5l}
\end{equation}%
\begin{equation}
m_{13i}=(v_{ix}^{2}+v_{iy}^{2})(v_{ix}^{2}-v_{iy}^{2})\text{,}  \label{5m}
\end{equation}%
\begin{equation}
m_{14i}=(v_{ix}^{2}+v_{iy}^{2})v_{ix}v_{iy}\text{,}  \label{5n}
\end{equation}%
\begin{equation}
m_{15i}=v_{ix}(v_{ix}^{2}-3v_{iy}^{2})(v_{ix}^{2}+v_{iy}^{2})\text{,}
\label{5o}
\end{equation}%
\begin{equation}
m_{16i}=v_{iy}(3v_{ix}^{2}-v_{iy}^{2})(v_{ix}^{2}+v_{iy}^{2})\text{.}
\label{5p}
\end{equation}%
where $i=1,\cdots ,16$.

For two-dimensional compressible models, we have four conserved moments,
density $\rho $, momentums $j_x$, $j_y$, and energy $e$. They are denoted by
$\hat{f}_{1}$, $\hat{f}_{2}$, $\hat{f}_{3}$ and $\hat{f}_{4}$, respectively.
Specifically, $\hat{f}_{1}=\rho =\sum f_{i}m_{1i}$, $\hat{f}_{2}=j_{x}=\sum
f_{i}m_{2i}$, $\hat{f}_{3}=j_{y}=\sum f_{i}m_{3i}$, $\hat{f}_{4}=e=\sum
f_{i}m_{4i}$. To be consistent with the idiomatic expression of energy, in
the definitions of $m_{4i}$, $m_{7i}$ and $m_{8i}$, a coefficient $1/2$ is
used. Similarly, a coefficient $1/4$ is used in the definition of $m_{11i}$.
Thus, the transformation matrix $\mathbf{M}$\ can be expressed as follows:
\end{subequations}
\begin{equation*}
\mathbf{M}=(m_{1},m_{2},\cdots ,m_{16})^{T}\text{,}
\end{equation*}%
where%
\begin{equation*}
m_{1}=(1,1,1,1,1,1,1,1,1,1,1,1,1,1,1,1)\text{,}
\end{equation*}%
\begin{equation*}
m_{2}=(1,0,-1,0,6,0,-6,0,\sqrt{2},-\sqrt{2},-\sqrt{2},\sqrt{2},\frac{3}{%
\sqrt{2}},-\frac{3}{\sqrt{2}},-\frac{3}{\sqrt{2}},\frac{3}{\sqrt{2}})\text{,}
\end{equation*}%
\begin{equation*}
m_{3}=(0,1,0,-1,0,6,0,-6,\sqrt{2},\sqrt{2},-\sqrt{2},-\sqrt{2},\frac{3}{%
\sqrt{2}},\frac{3}{\sqrt{2}},-\frac{3}{\sqrt{2}},-\frac{3}{\sqrt{2}})\text{,}
\end{equation*}%
\begin{equation*}
m_{4}=(\frac{1}{2},\frac{1}{2},\frac{1}{2},\frac{1}{2},18,18,18,18,2,2,2,2,%
\frac{9}{2},\frac{9}{2},\frac{9}{2},\frac{9}{2})\text{,}
\end{equation*}%
\begin{equation*}
m_{5}=(1,-1,1,-1,36,-36,36,-36,0,0,0,0,0,0,0,0)\text{,}
\end{equation*}%
\begin{equation*}
m_{6}=(0,0,0,0,0,0,0,0,2,-2,2,-2,\frac{9}{2},-\frac{9}{2},\frac{9}{2},-\frac{%
9}{2})\text{,}
\end{equation*}%
\begin{equation*}
m_{7}=(\frac{1}{2},0,-\frac{1}{2},0,108,0,-108,0,2\sqrt{2},-2\sqrt{2},-2%
\sqrt{2},2\sqrt{2},\frac{27}{2\sqrt{2}},-\frac{27}{2\sqrt{2}},-\frac{27}{2%
\sqrt{2}},\frac{27}{2\sqrt{2}})\text{,}
\end{equation*}%
\begin{equation*}
m_{8}=(0,\frac{1}{2},0,-\frac{1}{2},0,108,0,-108,2\sqrt{2},2\sqrt{2},-2\sqrt{%
2},-2\sqrt{2},\frac{27}{2\sqrt{2}},\frac{27}{2\sqrt{2}},-\frac{27}{2\sqrt{2}}%
,-\frac{27}{2\sqrt{2}})\text{,}
\end{equation*}%
\begin{equation*}
m_{9}=(1,0,-1,0,216,0,-216,0,-4\sqrt{2},4\sqrt{2},4\sqrt{2},-4\sqrt{2},-%
\frac{27}{\sqrt{2}},\frac{27}{\sqrt{2}},\frac{27}{\sqrt{2}},-\frac{27}{\sqrt{%
2}})\text{,}
\end{equation*}%
\begin{equation*}
m_{10}=(0,-1,0,1,0,-216,0,216,4\sqrt{2},4\sqrt{2},-4\sqrt{2},-4\sqrt{2},%
\frac{27}{\sqrt{2}},\frac{27}{\sqrt{2}},-\frac{27}{\sqrt{2}},-\frac{27}{%
\sqrt{2}})\text{,}
\end{equation*}%
\begin{equation*}
m_{11}=(\frac{1}{4},\frac{1}{4},\frac{1}{4},\frac{1}{4}%
,324,324,324,324,4,4,4,4,\frac{81}{4},\frac{81}{4},\frac{81}{4},\frac{81}{4})%
\text{,}
\end{equation*}%
\begin{equation*}
m_{12}=(1,1,1,1,1296,1296,1296,1296,-16,-16,-16,-16,-81,-81,-81,-81)\text{,}
\end{equation*}%
\begin{equation*}
m_{13}=(1,-1,1,-1,1296,-1296,1296,-1296,0,0,0,0,0,0,0,0)\text{,}
\end{equation*}%
\begin{equation*}
m_{14}=(0,0,0,0,0,0,0,0,8,-8,8,-8,\frac{81}{2},-\frac{81}{2},\frac{81}{2},-%
\frac{81}{2})\text{,}
\end{equation*}%
\begin{equation*}
m_{15}=(1,0,-1,0,7776,0,-7776,0,-16\sqrt{2},16\sqrt{2},16\sqrt{2},-16\sqrt{2}%
,-\frac{243}{\sqrt{2}},\frac{243}{\sqrt{2}},\frac{243}{\sqrt{2}},-\frac{243}{%
\sqrt{2}})\text{,}
\end{equation*}%
\begin{equation*}
m_{16}=(0,-1,0,1,0,-7776,0,7776,16\sqrt{2},16\sqrt{2},-16\sqrt{2},-16\sqrt{2}%
,\frac{243}{\sqrt{2}},\frac{243}{\sqrt{2}},-\frac{243}{\sqrt{2}},-\frac{243}{%
\sqrt{2}})\text{.}
\end{equation*}

\subsection{Determination of $\hat{f}_{i}^{eq}$}

We perform the Chapman-Enskog expansion\cite{20,22,23} on the two sides of
Eq.\eqref{1}. We define
\begin{subequations}
\begin{equation}
f_{i}=f_{i}^{(0)}+f_{i}^{(1)}+f_{i}^{(2)}\text{,}  \label{6a}
\end{equation}%
\begin{equation}
\frac{\partial }{\partial t}=\frac{\partial }{\partial t_{1}}+\frac{\partial
}{\partial t_{2}}\text{,}  \label{6b}
\end{equation}%
\begin{equation}
\frac{\partial }{\partial x}=\frac{\partial }{\partial x_{1}}\text{,}
\label{6c}
\end{equation}%
where $f_{i}^{(0)}$\ is the zeroth order, $f_{i}^{(1)}$,\ $\frac{\partial }{%
\partial t_{1}}$and\ $\frac{\partial }{\partial x_{1}}$\ are the first
order, $f_{i}^{(2)}$\ and $\frac{\partial }{\partial t_{2}}$ are the second
order terms of the Knudsen number $\epsilon $. Equating the coefficients of
the zeroth, the first, and the second order terms in $\epsilon $ gives
\end{subequations}
\begin{subequations}
\begin{equation}
f_{i}^{(0)}=f_{i}^{eq}\text{,}  \label{7a}
\end{equation}%
\begin{equation}
(\frac{\partial }{\partial t_{1}}+v_{i\alpha }\frac{\partial }{\partial
x_{1\alpha }})f_{i}^{(0)}=-\mathbf{S}_{il}f_{l}^{(1)}\text{,}  \label{7b}
\end{equation}%
\begin{equation}
\frac{\partial }{\partial t_{2}}f_{i}^{(0)}+(\frac{\partial }{\partial t_{1}}%
+v_{i\alpha }\frac{\partial }{\partial x_{1\alpha }})f_{i}^{(1)}=-\mathbf{S}%
_{il}f_{l}^{(2)}\text{.}  \label{7c}
\end{equation}%
They can be converted into moment space to obtain:
\end{subequations}
\begin{subequations}
\begin{equation}
\hat{f}_{i}^{(0)}=\hat{f}_{i}^{eq}\text{,}  \label{8a}
\end{equation}%
\begin{equation}
(\frac{\partial }{\partial t_{1}}+\hat{\mathbf{E}}_{\alpha }\frac{\partial }{%
\partial x_{1\alpha }})\hat{f}_{i}^{(0)}=-\hat{\mathbf{S}}_{il}\hat{f}%
_{l}^{(1)}\text{,}  \label{8b}
\end{equation}%
\begin{equation}
\frac{\partial }{\partial t_{2}}\hat{f}_{i}^{(0)}+(\frac{\partial }{\partial
t_{1}}+\hat{\mathbf{E}}_{\alpha }\frac{\partial }{\partial x_{1\alpha }})%
\hat{f}_{i}^{(1)}=-\hat{\mathbf{S}}_{il}\hat{f}_{l}^{(2)}\text{,}  \label{8c}
\end{equation}%
where $\hat{\mathbf{E}}_{\alpha }=\mathbf{M}(v_{i\alpha }\mathbf{I})\mathbf{M%
}^{-1}$.

The equilibria of the moments in the moment space can be defined as : $\hat{%
\mathbf{f}}^{eq}=(\rho ,j_{x},j_{y},e,\hat{f}_{5}^{eq},\hat{f}%
_{6}^{eq},\cdots,\hat{f}_{16}^{eq})^{T}$. The first order deviations from
equilibria are defined as : $\hat{\mathbf{f}}^{(1)}=(0,0,0,0,\hat{f}%
_{5}^{(1)},\hat{f}_{6}^{(1)},\cdots,\hat{f}_{16}^{(1)})^{T}$. In the same
way, the second order deviations are $\hat{\mathbf{f}}^{(2)}=(0,0,0,0,\hat{f}%
_{5}^{(2)},\hat{f}_{6}^{(2)},\cdots,\hat{f}_{16}^{(2)})^{T}$. From Eq.%
\eqref{8b} we obtain
\end{subequations}
\begin{subequations}
\begin{equation}
\frac{\partial \rho }{\partial t_{1}}+\frac{\partial j_{x}}{\partial x_{1}}+%
\frac{\partial j_{y}}{\partial y_{1}}=0\text{,}  \label{9a}
\end{equation}%
\begin{equation}
\frac{\partial j_{x}}{\partial t_{1}}+\frac{\partial }{\partial x_{1}}(e+%
\frac{1}{2}\hat{f}_{5}^{eq})+\frac{\partial }{\partial y_{1}}\hat{f}%
_{6}^{eq}=0\text{,}  \label{9b}
\end{equation}%
\begin{equation}
\frac{\partial j_{y}}{\partial t_{1}}+\frac{\partial }{\partial x_{1}}\hat{f}%
_{6}^{eq}+\frac{\partial }{\partial y_{1}}(e-\frac{1}{2}\hat{f}_{5}^{eq})=0%
\text{,}  \label{9c}
\end{equation}%
\begin{equation}
\frac{\partial e}{\partial t_{1}}+\frac{\partial }{\partial x_{1}}\hat{f}%
_{7}^{eq}+\frac{\partial }{\partial y_{1}}\hat{f}_{8}^{eq}=0\text{,}
\label{9d}
\end{equation}%
\begin{equation}
\frac{\partial }{\partial t_{1}}\hat{f}_{5}^{eq}+\frac{\partial }{\partial
x_{1}}(\hat{f}_{7}^{eq}+\frac{1}{2}\hat{f}_{9}^{eq})+\frac{\partial }{%
\partial y_{1}}(-\hat{f}_{8}^{eq}+\frac{1}{2}\hat{f}_{10}^{eq})=-s_{5}\hat{f}%
_{5}^{(1)}\text{,}  \label{9e}
\end{equation}%
\begin{equation}
\frac{\partial }{\partial t_{1}}\hat{f}_{6}^{eq}+\frac{1}{4}\frac{\partial }{%
\partial x_{1}}(2\hat{f}_{8}^{eq}+\hat{f}_{10}^{eq})+\frac{1}{4}\frac{%
\partial }{\partial y_{1}}(2\hat{f}_{7}^{eq}-\hat{f}_{9}^{eq})=-s_{6}\hat{f}%
_{6}^{(1)}\text{,}  \label{9f}
\end{equation}%
\begin{equation}
\frac{\partial }{\partial t_{1}}\hat{f}_{7}^{eq}+\frac{\partial }{\partial
x_{1}}(\hat{f}_{11}^{eq}+\frac{1}{4}\hat{f}_{13}^{eq})+\frac{1}{2}\frac{%
\partial }{\partial y_{1}}\hat{f}_{14}^{eq}=-s_{7}\hat{f}_{7}^{(1)}\text{,}
\label{9g}
\end{equation}%
\begin{equation}
\frac{\partial }{\partial t_{1}}\hat{f}_{8}^{eq}+\frac{1}{2}\frac{\partial }{%
\partial x_{1}}\hat{f}_{14}^{eq}+\frac{\partial }{\partial y_{1}}(\hat{f}%
_{11}^{eq}-\frac{1}{4}\hat{f}_{13}^{eq})=-s_{8}\hat{f}_{8}^{(1)}\text{,}
\label{9h}
\end{equation}%
\begin{equation}
\frac{\partial }{\partial t_{1}}\hat{f}_{9}^{eq}+\frac{1}{2}\frac{\partial }{%
\partial x_{1}}(\hat{f}_{12}^{eq}+\hat{f}_{13}^{eq})-\frac{\partial }{%
\partial y_{1}}\hat{f}_{14}^{eq}=-s_{9}\hat{f}_{9}^{(1)}\text{,}  \label{9i}
\end{equation}%
\begin{equation}
\frac{\partial }{\partial t_{1}}\hat{f}_{10}^{eq}+\frac{\partial }{\partial
x_{1}}\hat{f}_{14}^{eq}+\frac{1}{2}\frac{\partial }{\partial y_{1}}(-\hat{f}%
_{12}^{eq}+\hat{f}_{13}^{eq})=-s_{10}\hat{f}_{10}^{(1)}\text{,}  \label{9j}
\end{equation}%
\begin{equation}
\frac{\partial }{\partial t_{1}}\hat{f}_{11}^{eq}+\frac{\partial }{\partial
x_{1}}(-9j_{x}+\frac{25}{2}\hat{f}_{7}^{eq}+3\hat{f}_{9}^{eq})+\frac{%
\partial }{\partial y_{1}}(-9j_{y}+\frac{25}{2}\hat{f}_{8}^{eq}-3\hat{f}%
_{10}^{eq})=-s_{11}\hat{f}_{11}^{(1)}\text{,}  \label{9k}
\end{equation}%
\begin{equation}
\frac{\partial }{\partial t_{1}}\hat{f}_{12}^{eq}+\frac{\partial }{\partial
x_{1}}\hat{f}_{15}^{eq}-\frac{\partial }{\partial y_{1}}\hat{f}%
_{16}^{eq}=-s_{12}\hat{f}_{12}^{(1)}\text{,}  \label{9l}
\end{equation}%
\begin{equation}
\frac{\partial }{\partial t_{1}}\hat{f}_{13}^{eq}+\frac{\partial }{\partial
x_{1}}(-18j_{x}+\frac{1}{2}\hat{f}_{15}^{eq}+25\hat{f}_{7}^{eq}+6\hat{f}%
_{9}^{eq})+\frac{\partial }{\partial y_{1}}(18j_{y}+\frac{1}{2}\hat{f}%
_{16}^{eq}-25\hat{f}_{8}^{eq}+6\hat{f}_{10}^{eq})=-s_{13}\hat{f}_{13}^{(1)}%
\text{,}  \label{9m}
\end{equation}%
\begin{equation}
\frac{\partial }{\partial t_{1}}\hat{f}_{14}^{eq}+\frac{\partial }{\partial
x_{1}}(-9j_{y}+\frac{1}{4}\hat{f}_{16}^{eq}+\frac{25}{2}\hat{f}_{8}^{eq}-3%
\hat{f}_{10}^{eq})-\frac{\partial }{\partial y_{1}}(9j_{x}+\frac{1}{4}\hat{f}%
_{15}^{eq}-\frac{25}{2}\hat{f}_{7}^{eq}-3\hat{f}_{9}^{eq})=-s_{14}\hat{f}%
_{14}^{(1)}\text{,}  \label{9n}
\end{equation}%
\begin{equation}
\frac{\partial }{\partial t_{1}}\hat{f}_{15}^{eq}+\frac{\partial }{\partial
x_{1}}(75e-54\rho -18\hat{f}_{5}^{eq}+18\hat{f}_{11}^{eq}+\frac{25}{2}\hat{f}%
_{12}^{eq}+\frac{37}{2}\hat{f}_{13}^{eq})+\frac{\partial }{\partial y_{1}}(36%
\hat{f}_{6}^{eq}-13\hat{f}_{14}^{eq})=-s_{15}\hat{f}_{15}^{(1)}\text{,}
\label{9o}
\end{equation}%
\begin{equation}
\frac{\partial }{\partial t_{1}}\hat{f}_{16}^{eq}-\frac{\partial }{\partial
x_{1}}(36\hat{f}_{6}^{eq}-13\hat{f}_{14}^{eq})-\frac{\partial }{\partial
y_{1}}(75e-54\rho +18\hat{f}_{5}^{eq}+18\hat{f}_{11}^{eq}+\frac{25}{2}\hat{f}%
_{12}^{eq}-\frac{37}{2}\hat{f}_{13}^{eq})=-s_{16}\hat{f}_{16}^{(1)}\text{.}
\label{9p}
\end{equation}%
From Eq.\eqref{8c} we obtain
\end{subequations}
\begin{subequations}
\begin{equation}
\frac{\partial \rho }{\partial t_{2}}=0\text{,}  \label{10a}
\end{equation}%
\begin{equation}
\frac{\partial j_{x}}{\partial t_{2}}+\frac{1}{2}\frac{\partial }{\partial
x_{1}}\hat{f}_{5}^{(1)}+\frac{\partial }{\partial y_{1}}\hat{f}_{6}^{(1)}=0%
\text{,}  \label{10b}
\end{equation}%
\begin{equation}
\frac{\partial j_{y}}{\partial t_{2}}+\frac{\partial }{\partial x_{1}}\hat{f}%
_{6}^{(1)}-\frac{1}{2}\frac{\partial }{\partial y_{1}}\hat{f}_{5}^{(1)}=0%
\text{,}  \label{10c}
\end{equation}%
\begin{equation}
\frac{\partial e}{\partial t_{2}}+\frac{\partial }{\partial x_{1}}\hat{f}%
_{7}^{(1)}+\frac{\partial }{\partial y_{1}}\hat{f}_{8}^{(1)}=0\text{.}
\label{10d}
\end{equation}%
Adding Eq.(10) and the first four formulas of Eq.(9) leads to the following
equations,
\end{subequations}
\begin{subequations}
\begin{equation}
\frac{\partial \rho }{\partial t}+\frac{\partial j_{x}}{\partial x}+\frac{%
\partial j_{y}}{\partial y}=0\text{,}  \label{11a}
\end{equation}%
\begin{equation}
\frac{\partial j_{x}}{\partial t}+\frac{\partial }{\partial x}(e+\frac{1}{2}%
\hat{f}_{5}^{eq})+\frac{\partial }{\partial y}\hat{f}_{6}^{eq}=-\frac{1}{2}%
\frac{\partial }{\partial x}\hat{f}_{5}^{(1)}-\frac{\partial }{\partial y}%
\hat{f}_{6}^{(1)}\text{,}  \label{11b}
\end{equation}%
\begin{equation}
\frac{\partial j_{y}}{\partial t}+\frac{\partial }{\partial x}\hat{f}%
_{6}^{eq}+\frac{\partial }{\partial y}(e-\frac{1}{2}\hat{f}_{5}^{eq})=-\frac{%
\partial }{\partial x}\hat{f}_{6}^{(1)}+\frac{1}{2}\frac{\partial }{\partial
y}\hat{f}_{5}^{(1)}\text{,}  \label{11c}
\end{equation}%
\begin{equation}
\frac{\partial e}{\partial t}+\frac{\partial }{\partial x}\hat{f}_{7}^{eq}+%
\frac{\partial }{\partial y}\hat{f}_{8}^{eq}=-\frac{\partial }{\partial x}%
\hat{f}_{7}^{(1)}-\frac{\partial }{\partial y}\hat{f}_{8}^{(1)}\text{.}
\label{11d}
\end{equation}

To obatin the NS equations, we choose
\end{subequations}
\begin{subequations}
\begin{equation}
\hat{f}_{5}^{eq}=(j_{x}^{2}-j_{y}^{2})/\rho \text{,}  \label{14a}
\end{equation}%
\begin{equation}
\hat{f}_{6}^{eq}=j_{x}j_{y}/\rho \text{,}  \label{14b}
\end{equation}%
\begin{equation}
\hat{f}_{7}^{eq}=(e+P)j_{x}/\rho \text{,}  \label{14c}
\end{equation}%
\begin{equation}
\hat{f}_{8}^{eq}=(e+P)j_{y}/\rho \text{,}  \label{14d}
\end{equation}%
\begin{equation}
\hat{f}_{9}^{eq}=(j_{x}^{2}-3j_{y}^{2})j_{x}/\rho ^{2}\text{,}  \label{14e}
\end{equation}%
\begin{equation}
\hat{f}_{10}^{eq}=(3j_{x}^{2}-j_{y}^{2})j_{y}/\rho ^{2}\text{,}  \label{14f}
\end{equation}%
\begin{equation}
\hat{f}_{11}^{eq}=2e^{2}/\rho -(j_{x}^{2}+j_{y}^{2})^{2}/4\rho ^{3}\text{,}
\label{14g}
\end{equation}%
\begin{equation}
\hat{f}_{13}^{eq}=(6\rho e-2j_{x}^{2}-2j_{y}^{2})(j_{x}^{2}-j_{y}^{2})/\rho
^{3}\text{,}  \label{14h}
\end{equation}%
\begin{equation}
\hat{f}_{14}^{eq}=(6\rho e-2j_{x}^{2}-2j_{y}^{2})j_{x}j_{y}/\rho ^{3}\text{.}
\label{14i}
\end{equation}%
The definitions of $\hat{f}_{12}^{eq}$,\ $\hat{f}_{15}^{eq}$,\ $\hat{f}%
_{16}^{eq}$ have no effect on macroscopic equations, so we can choose $\hat{f%
}_{12}^{eq}=\hat{f}_{15}^{eq}=\hat{f}_{16}^{eq}=0$. In this way the
recovered NS equations are as follows:
\end{subequations}
\begin{subequations}
\begin{equation}
\frac{\partial \rho }{\partial t}+\frac{\partial j_{x}}{\partial x}+\frac{%
\partial j_{y}}{\partial y}=0\text{,}  \label{15a}
\end{equation}%
\begin{equation}
\frac{\partial j_{x}}{\partial t}+\frac{\partial }{\partial x}\left(
j_{x}^{2}/\rho \right) +\frac{\partial }{\partial y}\left( j_{x}j_{y}/\rho
\right) =-\frac{\partial P}{\partial x}+\frac{\partial }{\partial x}[\mu
_{s}(\frac{\partial u_{x}}{\partial x}-\frac{\partial u_{y}}{\partial y})]+%
\frac{\partial }{\partial y}[\mu _{v}(\frac{\partial u_{y}}{\partial x}+%
\frac{\partial u_{x}}{\partial y})]\text{,}  \label{15b}
\end{equation}%
\begin{equation}
\frac{\partial j_{y}}{\partial t}+\frac{\partial }{\partial x}\left(
j_{x}j_{y}/\rho \right) +\frac{\partial }{\partial y}\left( j_{y}^{2}/\rho
\right) =-\frac{\partial P}{\partial y}+\frac{\partial }{\partial x}[\mu
_{v}(\frac{\partial u_{y}}{\partial x}+\frac{\partial u_{x}}{\partial y})]-%
\frac{\partial }{\partial y}[\mu _{s}(\frac{\partial u_{x}}{\partial x}-%
\frac{\partial u_{y}}{\partial y})]\text{,}  \label{15c}
\end{equation}%
\begin{eqnarray}
&&\frac{\partial e}{\partial t}+\frac{\partial }{\partial x}[(e+P)j_{x}/\rho
]+\frac{\partial }{\partial y}[(e+P)j_{y}/\rho ]  \notag \\
&=&\frac{\partial }{\partial x}[\lambda _{1}(2\frac{\partial T}{\partial x}%
+u_{y}\frac{\partial u_{y}}{\partial x}+u_{x}\frac{\partial u_{x}}{\partial x%
}-u_{x}\frac{\partial u_{y}}{\partial y}+u_{y}\frac{\partial u_{x}}{\partial
y})]  \notag \\
&&+\frac{\partial }{\partial y}[\lambda _{2}(2\frac{\partial T}{\partial y}%
+u_{x}\frac{\partial u_{x}}{\partial y}-u_{y}\frac{\partial u_{x}}{\partial x%
}+u_{x}\frac{\partial u_{y}}{\partial x}+u_{y}\frac{\partial u_{y}}{\partial
y})]\text{,}  \label{15d}
\end{eqnarray}%
where $\mu _{s}=$ $\rho RT/s_{5}$, $\mu _{v}=$ $\rho RT/s_{6}$, $\lambda
_{1}=\rho RT/s_{7}$, $\lambda _{2}=\rho RT/s_{8}$. When $\mu _{s}=$\ $\mu
_{v}=\mu $,\ $\lambda _{1}=\lambda _{2}=\lambda $, the above NS equations
reduce to
\end{subequations}
\begin{subequations}
\begin{equation}
\frac{\partial \rho }{\partial t}+\frac{\partial j_{\alpha }}{\partial
x_{\alpha }}=0\text{,}  \label{16a}
\end{equation}%
\begin{equation}
\frac{\partial j_{\alpha }}{\partial t}+\frac{\partial \left( j_{\alpha
}j_{\beta }/\rho \right) }{\partial x_{\beta }}=-\frac{\partial P}{\partial
x_{\alpha }}+\frac{\partial }{\partial x_{\beta }}[\mu (\frac{\partial
u_{\alpha }}{\partial x_{\beta }}+\frac{\partial u_{\beta }}{\partial
x_{\alpha }}-\frac{\partial u_{\chi }}{\partial x_{\chi }}\delta _{\alpha
\beta })]\text{,}  \label{16b}
\end{equation}%
\begin{equation}
\frac{\partial e}{\partial t}+\frac{\partial }{\partial x_{\alpha }}%
[(e+P)j_{\alpha }/\rho ]=\frac{\partial }{\partial x_{\alpha }}[\lambda (2%
\frac{\partial T}{\partial x_{\alpha }}+u_{\beta }(\frac{\partial u_{\alpha }%
}{\partial x_{\beta }}+\frac{\partial u_{\beta }}{\partial x_{\alpha }}-%
\frac{\partial u_{\chi }}{\partial x_{\chi }}\delta _{\alpha \beta })]\text{.%
}  \label{16c}
\end{equation}


\section{Von Neumann Stability Analysis}

In this section we perform the von Neumann stability analysis on the new MRT
LB model. In the stability analysis, we write the solution of FD LB equation
in the form of Fourier series. If all the eigenvalues of the coefficient
matrix are less than 1, the algorithm is stable.

The distribution function $f_{i}(x_{\alpha },t)$ is split into two parts:
\end{subequations}
\begin{equation}
f_{i}(x_{\alpha },t)=\bar{f_{i}^{0}}+\Delta f_{i}(x_{\alpha },t)\text{,}
\label{global}
\end{equation}%
where $\bar{f_{i}^{0}}$\ is the global equilibrium distribution function. It
is a constant and does not vary with time or space, depends only on the
average density, velocity and temperature. Similarly, the distribution
function $\hat{f}_{i}(x_{\alpha },t)$ is split into two parts:
\begin{equation}
\hat{f}_{i}(x_{\alpha },t)=\hat{\bar{f_{i}^{0}}}+\Delta \hat{f}%
_{i}(x_{\alpha },t)\text{,}  \label{mg}
\end{equation}%
where $\hat{\bar{f_{i}^{0}}}$ is a constant. Putting the Eq.%
\eqref{global} and Eq.\eqref{mg} into Eq.\eqref{3} gives
\begin{equation}
\Delta f_{i}(x_{\alpha },t+\Delta t)=\Delta f_{i}(x_{\alpha },t)-\Delta
tv_{i\alpha }\frac{\partial \Delta f_{i}}{\partial x_{\alpha }}-\Delta t%
\mathbf{M}_{il}^{-1}\hat{\mathbf{S}}_{lk}(\Delta \hat{f}_{k}-\Delta \hat{f}%
_{k}^{eq})\text{.}  \label{fi1}
\end{equation}%
The solution can be written as%
\begin{equation}
\Delta f_{i}(x_{\alpha },t)=F_{i}^{t}\mathrm{exp}(\mathbf{i}k_{\alpha
}x_{\alpha })\text{,}  \label{fi1jie}
\end{equation}%
where $F_{i}^{t}$ is an amplitude of sine wave at lattice point $x_{\alpha }$
and time $t$, $k_{\alpha }$ is the wave number. From the above two equations
we can obtain $F_{i}^{t+\Delta t}=G_{ij}F_{j}^{t}.$ Coefficient matrix $%
G_{ij}$\ describes the growth rate of amplitude $F_{i}^{t}$\ in each time
step $\Delta t$. If $\omega $ denotes the eigenvalue of coefficient matrix $%
G_{ij}$, the von Neumann stability condition is $\mathrm{max}|\omega |\leq 1$%
. Coefficient matrix $G_{ij}$ can be expressed as follows,
\begin{align}
G_{ij}& =\delta _{ij}-\frac{v_{i\alpha }\Delta t}{2\Delta x_{\alpha }}(e^{%
\mathbf{i}k_{\alpha }\Delta x_{\alpha }}-e^{-\mathbf{i}k_{\alpha }\Delta
x_{\alpha }})\delta _{ij}+\frac{1}{2}(\frac{v_{i\alpha }\Delta t}{\Delta
x_{\alpha }})^{2}(e^{\mathbf{i}k_{\alpha }\Delta x_{\alpha }}-2  \notag \\
& +e^{-\mathbf{i}k_{\alpha }\Delta x_{\alpha }})\delta _{ij}-\Delta t\mathbf{%
M}_{il}^{-1}\hat{\mathbf{S}}_{lk}(\frac{\partial \hat{f}_{k}}{\partial f_{j}}%
-\frac{\partial \hat{f}_{k}^{eq}}{\partial f_{j}})\text{,}
\end{align}%
where
\begin{equation*}
\hat{f}_{k}=\mathbf{M}_{kp}f_{p}\text{,}
\end{equation*}%
\begin{equation}
\frac{\partial \hat{f}_{k}^{eq}}{\partial f_{j}}=\frac{\partial \hat{f}%
_{k}^{eq}}{\partial \rho }\frac{\partial \rho }{\partial f_{j}}+\frac{%
\partial \hat{f}_{k}^{eq}}{\partial T}\frac{\partial T}{\partial f_{j}}+%
\frac{\partial \hat{f}_{k}^{eq}}{\partial u_{\alpha }}\frac{\partial
u_{\alpha }}{\partial f_{j}}\text{.}
\end{equation}

Coefficient matrix $G_{ij}$ contains a large number of matrix elements (as
many as $16\times 16$), and every matrix element correlates with the
macroscopic quantities and model parameters, so analytic analysis is very
difficult. We conduct a quantitative analysis using Mathematica software.

In Fig.2 we show an example of stability comparison for the new MRT model
and its SRT version. The abscissa is for $kdx$, and the vertical axis is for
$|\omega |_{max}$\ which is the largest eigenvalue of coefficient matrix $%
G_{ij}$. The macroscopic values in stability analysis are chosen as follows:
$(\rho ,u_{1},u_{2},T)$ = $(2.0,10.0,0.0,2.0)$, other common parameters are:
$dx=dy=2\times 10^{-3}$, $dt=10^{-5}$. The relaxation time in SRT is $\tau
=10^{-5}$, while the collision parameters in MRT are $s_{5}=6500$, $%
s_{7}=s_{8}=9\times 10^{4}$, $s_{9}=8\times 10^{4}$, $s_{13}=7\times 10^{4}$%
, $s_{14}=8\times 10^{3}$, $s_{15}=2.5\times 10^{4}$, the others are $10^{5}$%
. In this case, the MRT scheme is stable, while the SRT version is not. It
is clear that, by choosing appropriate collision parameters, the stability
of MRT can be much better than the SRT.
\begin{figure}[tbp]
\center\includegraphics*[width=0.67\textwidth]{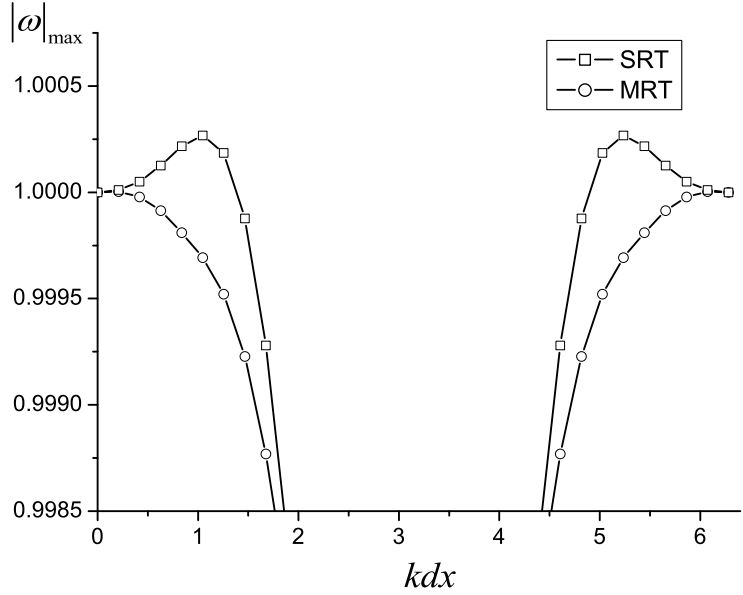}
\caption{Stability comparison for the new MRT model and its SRT version.}
\end{figure}

\section{Numerical Simulations}

In this section we study the following problems using the new MRT LB
model: One-dimensional Riemann problems, shock reflections, shock
wave reaction on cylindrical bubble, and Couette flow.

\subsection{One-dimensional Riemann problems}

Here, we study several typical one-dimensional Riemann problems,
including the Sod, Lax shock tube, Colella explosion wave, collision
of two strong shocks and a new shock tube with high Mach number. In
the $x$ direction, $f_{i}=$ $f_{i}^{eq}$ is set on the boundary
nodes before the disturbance reaches the two ends. In the $y$
direction, the periodic boundary condition is adopted. In the
following part, subscripts \textquotedblleft L\textquotedblright\
and \textquotedblleft R\textquotedblright\ indicate the macroscopic
variables at the left and right sides of discontinuity.

(a) Sod shock tube problem

The initial condition is
\begin{equation}
\left\{
\begin{array}{cc}
(\rho ,u_{1},u_{2},T)|_{L}=(1.0,0.0,0.0,1.0)\text{,} & x\leq 0\text{.} \\
(\rho ,u_{1},u_{2},T)|_{R}=(0.125,0.0,0.0,0.8)\text{,} & x>0\text{.}%
\end{array}%
\right.
\end{equation}

Figure 3 shows a comparison of the MRT LB simulation results and exact
solutions for the density, pressure, velocity and temperature of the Sod
shock tube problem at time $t=0.18$. Here, in the collision matrix $%
s_{5}=s_{6}=5\times 10^{2}$, $s_{7}=s_{8}=10^{3}$, $s_{11}=2500$,
and other collision parameters are$\ 10^{5}$. The red circles
correspond to simulation results with the grid size $dx=dy=0.002$
and time step $dt=2\times 10^{-6}$ (case 1), the green triangles
correspond to simulation results with $dx=dy=0.001$, $dt=10^{-6}$
(case 2), and solid lines represent the exact solutions. The
relative errors of the density, pressure, velocity and temperature
for case 1 are $0.234\%$, $0.182\%$, $3.32\%$ and $0.327\%$,
respectively. The relative errors of the density, pressure, velocity
and temperature for case 2 are $0.225\%$, $0.171\%$, $3.16\%$ and
$0.322\%$, respectively. Here, the relative error is defined as
$E=\sum_{I}\left\vert \varsigma _{_{I,J,num}}-\varsigma
_{_{I,J,exa}}\right\vert /\sum_{I}\left\vert \varsigma
_{_{I,J,exa}}\right\vert $, where $\varsigma _{_{I,J,num}}$ denotes
the variables at the node of $(x_{I},y_{J})$ obtained from the
numerical simulation, and $\varsigma _{_{I,J,exa}}$ is the exact
solution at the same node. The simulation results successfully
capture the main
structure of the waves. 
\begin{figure}[tbp]
\center\includegraphics*[width=0.76\textwidth]{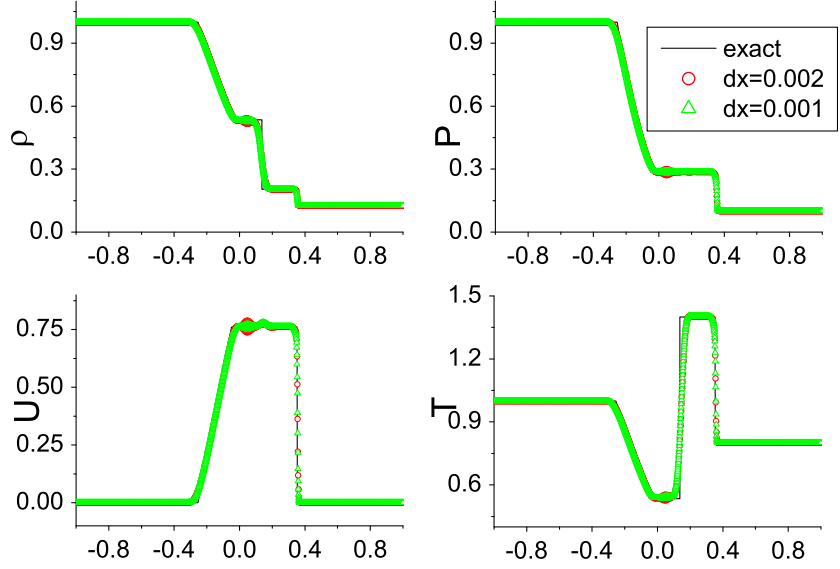}
\caption{Comparison of the MRT LB simulation results and exact solutions for
Sod shock tube at time $t=0.18$.}
\end{figure}

b) Lax shock tube problem

The initial condition of the problem is:
\begin{equation}
\left\{
\begin{array}{cc}
(\rho ,u_{1},u_{2},T)|_{L}=(0.445,0.698,0.0,7.928)\text{,} & x\leq 0\text{.}
\\
(\rho ,u_{1},u_{2},T)|_{R}=(0.5,0.0,0.0,1.142)\text{,} & x>0\text{.}%
\end{array}%
\right.
\end{equation}

Figure 4 shows the MRT LB numerical results and exact solutions for the Lax
shock tube problem at time $t=0.2$. The red squares, green circles and blue
triangles correspond to simulation results with different grid sizes and
time steps: $dx=dy=0.004$, $dt=4\times 10^{-6}$ (case 1), $dx=dy=0.002$, $%
dt=2\times 10^{-6}$ (case 2), $dx=dy=0.001$, $dt=10^{-6}$ (case 3),
respectively, and solid lines represent the exact solutions. The used
parameters are $s_{7}=s_{8}=3\times 10^{3}$, $s_{13}=10^{2}$, other
collision parameters are$\ 10^{5}$. The relative errors of the density,
pressure, velocity and temperature for case 1 are $0.398\%$, $0.205\%$, $%
0.592\%$ and $0.310\%$, respectively. The relative errors for case 2 are $%
0.344\%$, $0.130\%$, $0.408\%$ and $0.287\%$, respectively. And the relative
errors for case 3 are $0.334\%$, $0.117\%$, $0.372\%$ and $0.283\%$,
respectively. 
\begin{figure}[tbp]
\center\includegraphics*[width=0.76\textwidth]{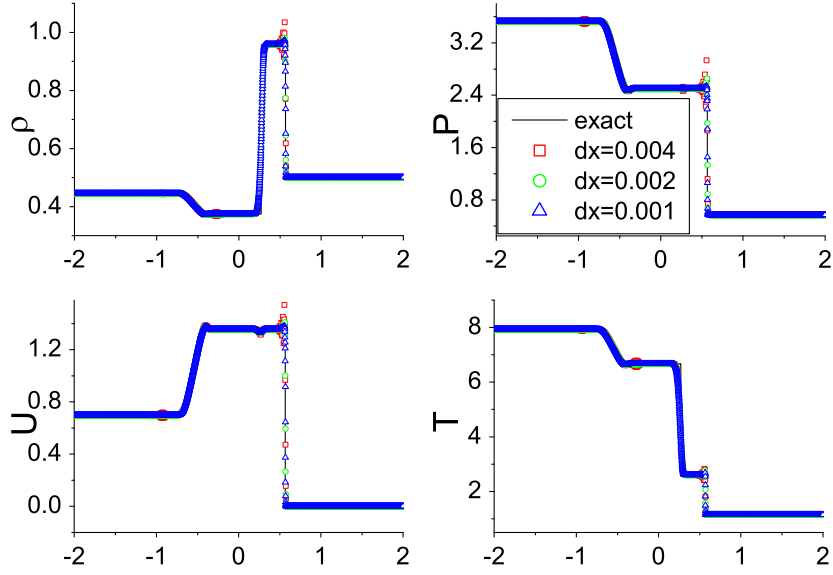}
\caption{The MRT LB numerical and exact solutions for Lax shock tube at time
$t=0.2$.}
\end{figure}

(c) Colella explosion wave

For the problem, the initial condition is
\begin{equation}
\left\{
\begin{array}{cc}
(\rho ,u_{1},u_{2},T)|_{L}=(1.0,0.0,0.0,1000.0)\text{,} & x\leq 0\text{.} \\
(\rho ,u_{1},u_{2},T)|_{R}=(1.0,0.0,0.0,0.01)\text{,} & x>0\text{.}%
\end{array}%
\right.
\end{equation}

This is a strong temperature discontinuity problem that can be used to study
the robustness and precision of numerical methods. Figure 5 gives density,
pressure, velocity and temperature results at $t=0.1$. The red squares and
green circles correspond to simulation results with different grid sizes and
time steps: $dx=dy=0.002$, $dt=2\times 10^{-6}$ (case 1), and $dx=dy=0.001$,
$dt=10^{-6}$ (case 2), respectively, and solid lines represent the exact
solutions. Here, the parameters are $s_{7}=s_{8}=5\times 10^{4}$, $%
s_{11}=s_{13}=5\times 10^{5}$, other values of $s$ adopt $10^{5}$.
The relative errors of the density, pressure, velocity and
temperature for case 1 are $1.69\%$, $1.11\%$, $1.60\% $ and
$0.779\%$, respectively. The relative errors for case 2 are
$1.68\%$, $1.11\%$, $1.59\%$ and $0.777\%$, respectively. The
oscillations at the interface are difficult to eliminate completely
in our simulations.
\begin{figure}[tbp]
\center\includegraphics*[width=0.76\textwidth]{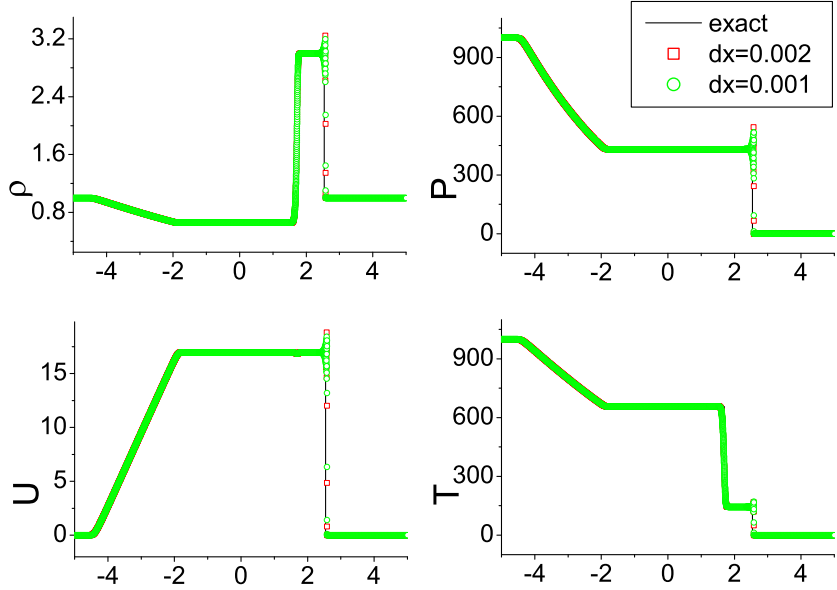}
\caption{The MRT simulation results and exact solutions for the Colella
explosion wave at time $t=0.1$.}
\end{figure}

(d) Collision of two strong shocks

The initial condition can be write as:
\begin{equation}
\left\{
\begin{array}{cc}
(\rho ,u_{1},u_{2},T)|_{L}=(5.99924,19.5975,0.0,76.8254)\text{,} & x\leq 0%
\text{.} \\
(\rho ,u_{1},u_{2},T)|_{R}=(5.99242,-6.19633,0.0,7.69222)\text{,} & x>0\text{%
.}%
\end{array}%
\right.
\end{equation}

The MRT and SRT numerical results and exact solutions at time $t=0.1$ are
shown in Fig.6, where the common parameters are $dx=dy=0.003$, $dt=10^{-5}$,
the collision matrix in MRT is $s_{5}=s_{6}=5\times 10^{3}$, $%
s_{7}=s_{8}=3\times 10^{4}$, other values of $s$\ are $10^{5}$, and the
relaxation time in SRT is $\tau =10^{-5}$. The red squares and green circles
correspond to the MRT and SRT simulation results, respectively, and solid
lines represent the exact solutions. Compared with the simulation results of
SRT, we can find that the oscillations at the discontinuity are weaker in
MRT simulation. 
\begin{figure}[tbp]
\center\includegraphics*[width=0.76\textwidth]{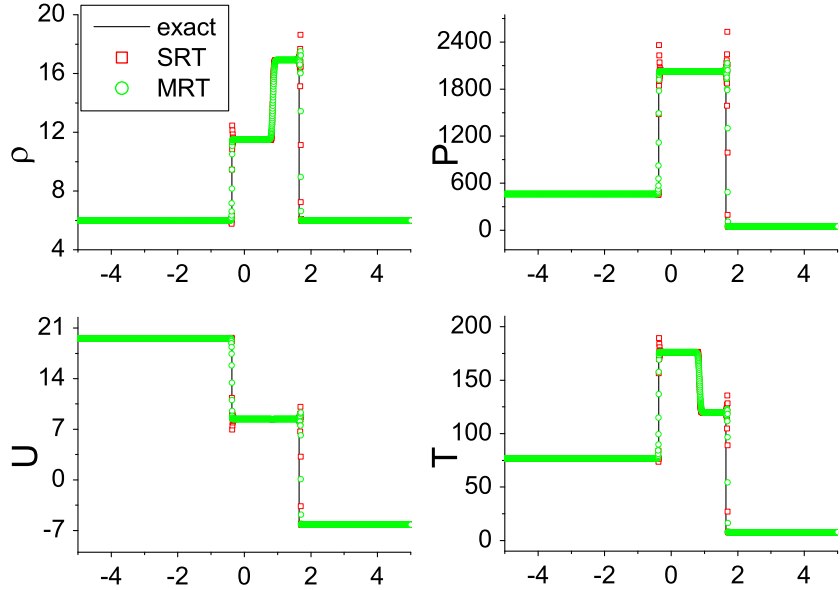}
\caption{The MRT and SRT numerical results and exact solution for collision
of two strong shocks at time $t=0.1$.}
\end{figure}

(e) High Mach number shock

In order to test the stability of the new model, we construct a new
shock tube problem with the initial condition,
\begin{equation}
\left\{
\begin{array}{cc}
(\rho ,u_{1},u_{2},T)|_{L}=(5.0,45.0,0.0,10.0)\text{,} & x\leq 0\text{.} \\
(\rho ,u_{1},u_{2},T)|_{R}=(6.0,-20.0,0.0,5.0)\text{,} & x>0\text{.}%
\end{array}%
\right.
\end{equation}%
The Mach number of the left side is $10.1$, and the right is $6.3$. And this
test is successfully passed by the MRT LB, but failed by the SRT. Figure 7
shows comparison of the MRT LB results and exact solutions at $t=0.018$,
where the parameters are $dx=dy=0.003$, $dt=10^{-5}$, $s_{5}=s_{6}=1.5\times
10^{4}$, $s_{10}=5\times 10^{4}$, other values of $s$\ are $10^{5}$. Circle
symbols correspond to MRT simulation results, and solid lines represent the
exact solutions. 
\begin{figure}[tbp]
\center\includegraphics*[width=0.76\textwidth]{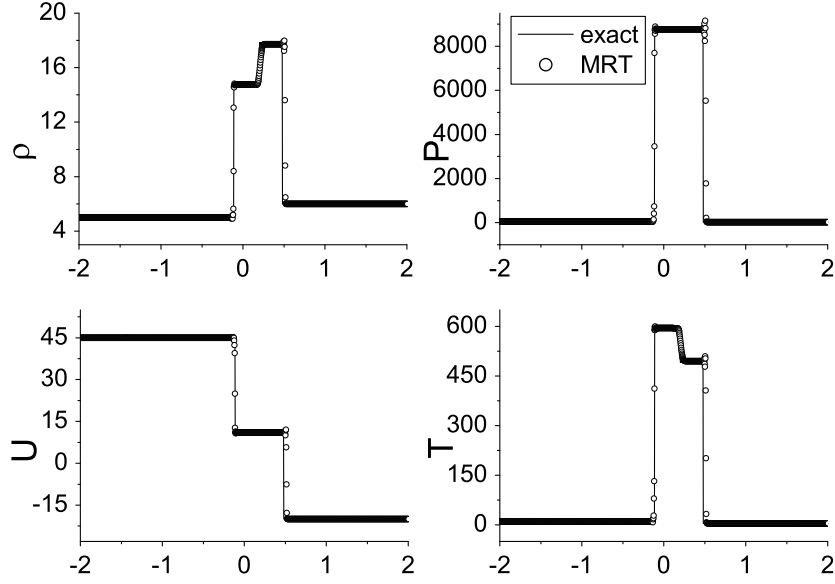}
\caption{MRT LB results and exact solutions for the high Mach number shock tube problem at $%
t=0.018$.}
\end{figure}

\subsection{Shock reflections}

We adopt the macroscopic variable boundary conditions in Figs. 8-11.
The values of the distribution functions on boundaries are assigned
with the corresponding values of the equilibrium distribution
functions. The determination methods of macroscopic quantities are
explained combining with specific problems. Here we study two gas
dynamic problems: regular and Mach shock reflections.

(a) Regular shock reflection

In the problem, we have performed a $30^{\circ }$\ shock reflection,
and the corresponding Mach number is $5$. The computational domain
is a rectangle with length $0.6$ and height $0.2$, which is divided
into $300\times 100.$ The reflecting wall lies at the bottom of the
domain (reflecting boundary condition denotes the $y$ component of
the fluid velocity on the boundary is reverse to that of interior
point), and the linear extrapolation technique is applied to define
the values of the macroscopic quantities on the right-hand boundary.
The other two sides adopt the Dirichlet boundary conditions:
\begin{equation}
\left\{
\begin{array}{l}
(\rho \text{, }u_{1}\text{, }u_{2}\text{, }T)|_{0\text{, }y\text{, }t}=(1.0%
\text{, }5.0\text{, }0.0\text{, }0.5)\text{,} \\
(\rho \text{, }u_{1}\text{, }u_{2}\text{, }T)|_{x\text{, }0.2\text{, }%
t}=(2.27273\text{, }4.3\text{, -1.21244, 1.76})\text{.}%
\end{array}%
\right.
\end{equation}%
Initially, the entire interior zone is set the values of the left boundary.
Parameters in the simulation are as follows: $dx=dy=0.002$, $dt=10^{-5}$, $%
s_{5}=10^{4}$, $s_{6}=2\times 10^{3}$, $s_{7}=s_{8}=10^{4}$, other collision
parameters are $10^{5}$. Figure 8 shows the pressure contour at time $t=0.3$%
, and the density, temperature contours have similar results. From black to
white, the grey level corresponds to the increase of the pressure. The
result shows that the new MRT model has the ability to accurately capture
the shock front. 
\begin{figure}[tbp]
\center\includegraphics*[width=0.6\textwidth]{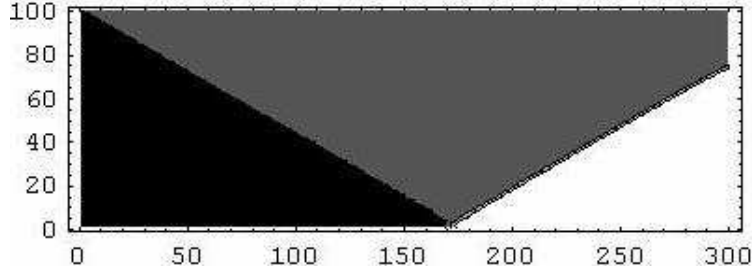}
\caption{Pressure contour of regular shock reflection at time $t=0.3$.}
\end{figure}

(b) Mach reflection problem

This problem is on an unsteady shock reflection. A planar shock impacts an
oblique surface which is at a $30^{\circ }$\ angle to the propagation
direction of the shock. The fluid in front of the shock has zero velocity,
and the shock Mach number is $1.5$. The initial condition is as follows:
\begin{equation}
\left( \rho ,u_{1},u_{2},T\right) \mid _{x\text{,}y\text{,}0}=\left\{
\begin{array}{ll}
(3.17647,0.555556\cos 30^{\circ },-0.555556\sin 30^{\circ },0.839506)\text{,}
& \text{ if }y\geq h(x,0)\text{.} \\
(2.0,0.0,0.0,0.5)\text{,} & \text{ if }y<h(x,0)\text{.}%
\end{array}%
\right.
\end{equation}%
where $h(x,t)=\tan 60^{\circ }(x-150dx)-1.5t/\sin 30^{\circ }.$ The
computational domain is divided into $600\times 300$. At the bottom
boundary, reflecting boundary condition is used from the 150th grid,
and the left side adopts the values of the initial post-shock flow;
the left boundary is also assigned values of the initial post-shock
flow, and at the right boundary the extrapolation technique is
applied; at the top boundary, the macroscopic variables are assigned
using the same method of the right boundary when $x>g(t)$, and are
set the same values as the left boundary's when $x\leq g(t)$, where
$g(t)=150dx+\tan 30^{\circ }(0.9+1.5t/\sin 30^{\circ })$, $dx$\ is
the grid size in simulation. In Fig.9 we show the results of
density, pressure, velocity in $x$\ direction and temperature
contours at $t=0.25$ in the part of $\left[ 50,450\right] \times \left[ 0,200%
\right] $. Parameters in the simulation are $dx=dy=0.003$, $dt=10^{-5}$, $%
s_{5}=10^{3}$, $s_{6}=5\times 10^{2}$, $s_{7}=s_{8}=10^{3}$, and other
collision parameters are$\ 10^{5}$. The simulation results are accordant
with those of other numerical methods\cite{41,42,43,CShuPRE}.
\begin{figure}[tbp]
\center\includegraphics*[width=0.9\textwidth]{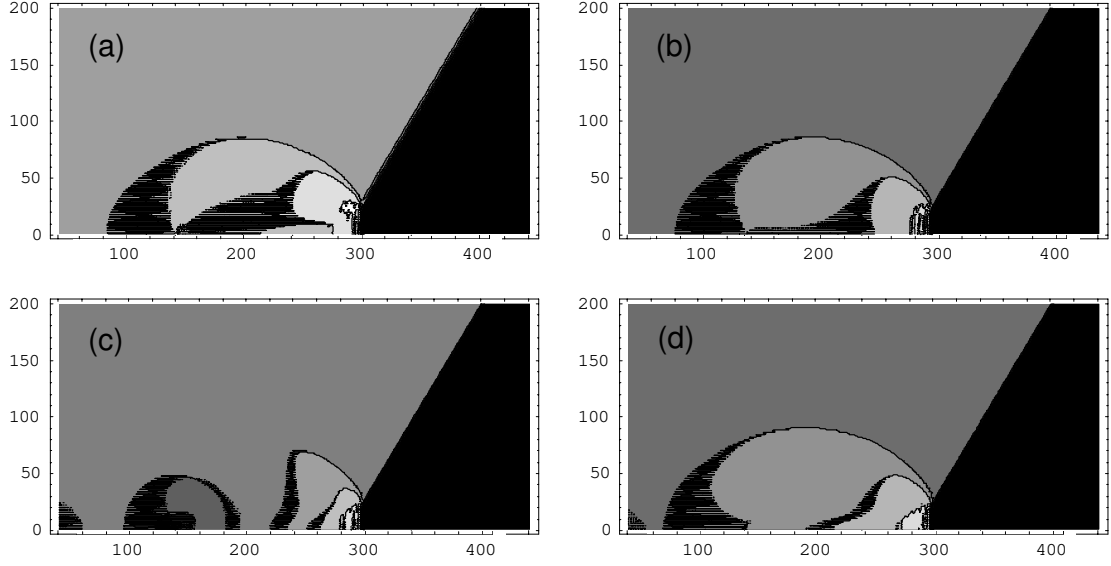}
\caption{Mach reflection of shock wave. Figs. (a)-(d) shows the density,
pressure, velocity in $x$ direction, and temperature at time $t=0.25$,
respectively. From black to white, the grey level corresponds to the
increase of values.}
\end{figure}

\subsection{Shock wave reaction on cylindrical bubble problem}

The problems are as follows: A planar shock wave with the Mach number $1.22$
impinges on a cylindrical bubble with different densities. In the first case
the bubble has a lower density. In the second case the bubble's density is
higher.

Initial conditions for the first case are
\begin{equation}
\left( \rho ,u_{1},u_{2},p\right) \mid _{x\text{,}y\text{,}0}=\left\{
\begin{array}{cc}
\left( 1,0,0,1\right) \text{,} & \mathbf{pre-shock}\text{\textbf{,}} \\
\left( 1.28,-0.3774,0,1.6512\right) \text{,} & \mathbf{post-shock}\text{%
\textbf{,}} \\
\left( 0.1358,0,0,1\right) \text{,} & \mathbf{bubble}\text{\textbf{,}}%
\end{array}%
\right.  \label{bubble1}
\end{equation}%
and for the second case are
\begin{equation}
\left( \rho ,u_{1},u_{2},p\right) \mid _{x\text{,}y\text{,}0}=\left\{
\begin{array}{cc}
\left( 1,0,0,1\right) \text{,} & \mathbf{pre-shock}\text{\textbf{,}} \\
\left( 1.28,-0.3774,0,1.6512\right) \text{,} & \mathbf{post-shock}\text{%
\textbf{,}} \\
\left( 3.1538,0,0,1\right) \text{,} & \mathbf{bubble}\text{\textbf{.}}%
\end{array}%
\right.  \label{bubble2}
\end{equation}

In the simulations, the right side adopts the values of the initial
post-shock flow; the extrapolation technique is applied at the left
boundary, and reflecting boundary conditions are imposed on the top
and bottom. The common parameters are as follows: $dx=dy=0.003$,
$dt=10^{-5}$. When simulate
the low density cylindrical bubble, the collision parameters are $%
s_{5}=s_{6}=s_{7}=s_{8}=10^{4}$, and others are $10^{5}$; when simulate the
high density bubble, the collision parameters are $s_{5}=10^{3}$, and $%
s=10^{5}$\ for the others. In Fig. 10(a), from top to bottom, the three
plots show the density contours at the times $t=0$, $0.5$, $0.65$,
respectively. In Fig. 10(b), from top to bottom, the three plots show the
density contours at the times $t=0$, $0.6$, $0.9$, respectively. These
results are accordant with those from other methods\cite{24} and experiment%
\cite{24aa}. The surface of bubbles is comparatively smooth, which indicates
that the MRT modle has high accuracy and resolution.
\begin{figure}[tbp]
\center\includegraphics*[width=0.67\textwidth]{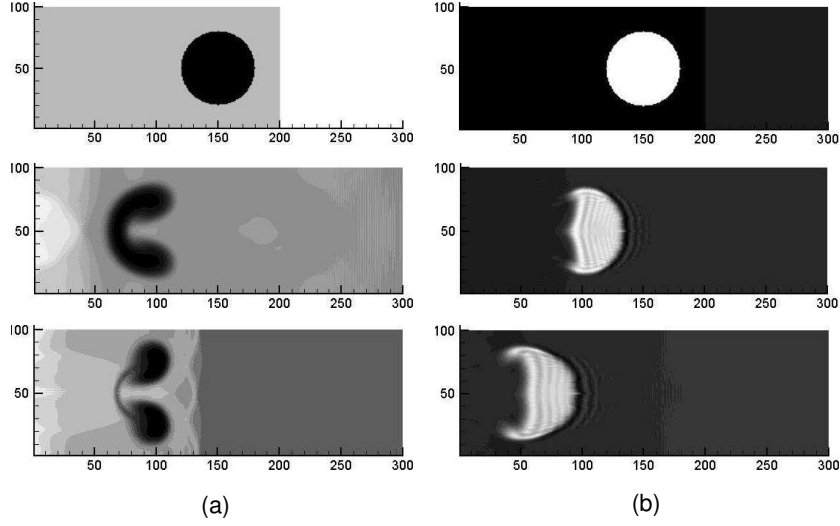}
\caption{Snapshots of shock wave reaction on single bubble. The left
column is for the process with initial condition \eqref{bubble1},
and the right column is for the process with initial condition
\eqref{bubble2}. From black to white, the density value increases.}
\end{figure}

\subsection{Couette flow}

In order to demonstrate the accuracy of the model, numerical
simulations of the incompressible Couette flow are carried out.
Consider a viscous fluid
flow between two infinite parallel flat plates, separated by a distance of $%
D $. The initial state of the fluid is $\rho =1$, $T=1$, $U=0$. At
time $t=0$ the two plates start to move at velocities $U$, $-U$,
respectively, ($U=0.1$). The periodic boundary condition is adopted
in the x direction. The top and bottom boundaries are constant speed
and constant temperature boundaries ($U=0.1, T=1$). The analytical
solution of horizontal velocity along a vertical line is as follows:
\begin{equation*}
u=2yU/D-\sum_{j}(-1)^{j+1}\frac{2U}{j\pi }\exp [-\frac{4j^{2}\pi ^{2}\mu }{%
\rho D^{2}}t]\sin (\frac{2j\pi }{D}y)\text{,}
\end{equation*}%
where $j$ is an integer, the two walls locate at $y=\pm D/2$.

We carried out a set of simulations: $dx=dy=0.004$, $%
dt=10^{-5}$, $NX\times NY=16\times 32$ (case 1), $dx=dy=0.002$,
$dt=5\times 10^{-6}$, $NX\times NY=32\times 64$(case 2),
$dx=dy=0.001$, $dt=2.5\times 10^{-6}$, $NX\times NY=64\times 128$
(case 3). All of the collision parameters are$\ 10^{5}$. Figure 11
shows a comparison of the MRT LB simulation results and exact
solution for the horizontal velocity distribution at time $t=57.5$.
The black squares, red triangles and green circles correspond to
simulation results of case 1, case 2, and case 3, respectively, and
solid line represents the exact solution.
The relative errors of the horizontal velocity for the three cases are $%
7.02\%$, $3.86\%$ and $2.06\%$, respectively. So this model is of first
order accuracy.%
\begin{figure}[tbp]
\center\includegraphics*[width=0.95\textwidth]{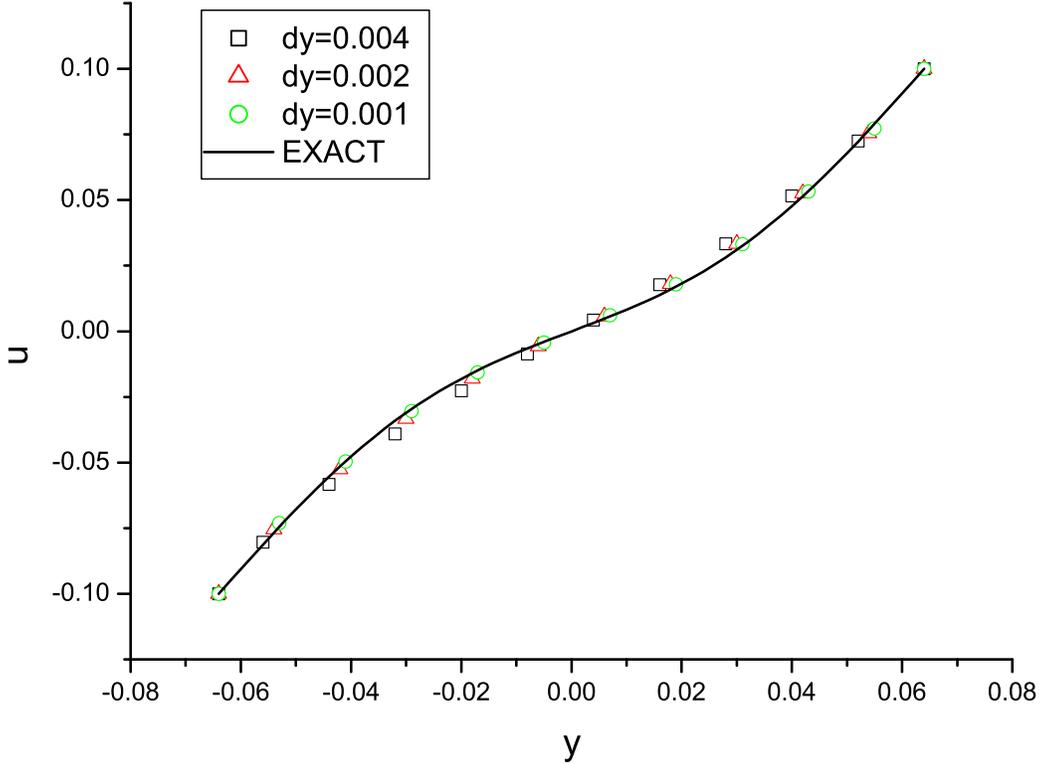}
\caption{Comparison of the MRT LB simulation results and exact solutions for
the horizontal velocity distribution of Couette fiow at time $t=57.5$.}
\end{figure}

\section{Conclusion}

An energy-conserving multiple-relaxation-time finite difference
lattice Boltzmann model for compressible flows is proposed. This
model is based on a 16-discrete-velocity model designed by Kataoka
and Tsutahara\cite{21a}. The collision step is first calculated in
the moment space and then mapped back to the velocity space. The
moment space and corresponding transformation matrix are constructed
according to the group representation theory. Equilibria of the
nonconserved moments are chosen according to the need of recovering
compressible Navier-Stokes equations through the Chapman-Enskog
expansion. In the new model different transport coefficients, such
as viscosity and heat conductivity, are related to different
collision parameters. Consequently, they can be controlled
independently. This flexibility makes the MRT model contain more
physical information and easier to satisfy the von Neumann stability
condition than its SRT counterpart. The new model passed well-known
benchmark tests, including (i) shock tubes, such as the Sod, Lax,
Colella explosion wave, collision of two strong shocks and a new
shock tube with high Mach number, (ii) regular and Mach shock
reflections, (iii) shock wave reaction on cylindrical bubble
problems, and (iv) Couette flow. This model works for both low and
high speeds compressible flows. Both the LB and traditional CFD
approach work when the Knudsen number is very small, but in the
vicinity of shock wave, the system is in a nonequilibrium state, and
the traditional Euler and Navier-Stokes descriptions are
problematic. For such problems, LB has more sound physical ground.


\section*{Acknowledgments}

The authors would like to thank Drs. Michael McCracken and Zhaoli Guo for
helpful discussions. This work is supported by the Science Foundations of
LCP and CAEP [under Grant Nos. 2009A0102005, 2009B0101012], National Basic
Research Program (973 Program) [under Grant No. 2007CB815105], National
Natural Science Foundation [under Grant Nos. 10775018, 10702010] of China.

\section*{Appendix. Spatial and temporal discretization effects}

In our simulations, the time evolution is based on the usual first-order
upwind scheme, while space discretization is performed through a
Lax-Wendroff scheme.
\begin{subequations}
\begin{equation}
\frac{\partial }{\partial t}f_{i}=[f_{i}(t+\delta t)-f_{i}(t)]/\delta t\text{%
,}  \label{A.1a}
\end{equation}%
\begin{eqnarray}
\overrightarrow{v}_{i}\cdot \nabla f_{i} &=&v_{ix}[f_{i}(x+\delta
x,y)-f_{i}(x-\delta x,y)]/2\delta x  \notag \\
&&+v_{iy}[f_{i}(x,y+\delta y)-f_{i}(x,y-\delta y)]/2\delta y  \notag \\
&&-v_{ix}^{2}\delta t[f_{i}(x+\delta x,y)-2f_{i}(x,y)+f_{i}(x-\delta
x,y)]/2\delta x^{2}  \notag \\
&&-v_{iy}^{2}\delta t[f_{i}(x,y+\delta y)-2f_{i}(x,y)+f_{i}(x,y-\delta
y)]/2\delta y^{2}\text{.}  \label{A.1b}
\end{eqnarray}%
If we perform the following series expansion up to second order in Eq.%
\eqref{A.1a} and Eq.\eqref{A.1b}
\end{subequations}
\begin{subequations}
\begin{equation}
f_{i}(t+\delta t)=f_{i}(t)+\delta t\frac{\partial }{\partial t}f_{i}+\frac{1%
}{2}\delta t^{2}\frac{\partial ^{2}}{\partial t^{2}}f_{i}\text{,}
\label{A.2a}
\end{equation}%
\begin{equation}
f_{i}(x+\delta x,y,t)=f_{i}(x,y,t)+\delta x\frac{\partial }{\partial x}f_{i}+%
\frac{1}{2}\delta x^{2}\frac{\partial ^{2}}{\partial x^{2}}f_{i}\text{,}
\label{A.2b}
\end{equation}%
\begin{equation}
f_{i}(x-\delta x,y,t)=f_{i}(x,y,t)-\delta x\frac{\partial }{\partial x}f_{i}+%
\frac{1}{2}\delta x^{2}\frac{\partial ^{2}}{\partial x^{2}}f_{i}\text{,}
\label{A.2c}
\end{equation}%
we can get
\end{subequations}
\begin{subequations}
\begin{equation}
\frac{\partial }{\partial t}f_{i}\Rightarrow \frac{\partial }{\partial t}%
f_{i}+\frac{1}{2}\delta t\frac{\partial ^{2}}{\partial t^{2}}f_{i}\text{,}
\label{A.3a}
\end{equation}%
\begin{equation}
\overrightarrow{v}_{i}\cdot \nabla f_{i}\Rightarrow v_{ix}\frac{\partial }{%
\partial x}f_{i}+v_{iy}\frac{\partial }{\partial y}f_{i}-\frac{%
v_{ix}^{2}\delta t}{2}\frac{\partial ^{2}}{\partial x^{2}}f_{i}-\frac{%
v_{iy}^{2}\delta t}{2}\frac{\partial ^{2}}{\partial y^{2}}f_{i}\text{,}
\label{A.3b}
\end{equation}%
So a more accurate LB equation solved using the updating schemes mentioned
above is,
\end{subequations}
\begin{equation}
\frac{\partial }{\partial t}f_{i}+\frac{1}{2}\delta t\frac{\partial ^{2}}{%
\partial t^{2}}f_{i}+v_{i\beta }\frac{\partial }{\partial r_{\beta }}f_{i}-%
\frac{1}{2}\delta tv_{i\beta }^{2}\frac{\partial ^{2}}{\partial r_{\beta
}^{2}}f_{i}=-\mathbf{S}_{ik}\left[ f_{k}-f_{k}^{eq}\right] \text{,}
\label{A.4}
\end{equation}

To investigate the effect of the supplementary terms in Eq.\eqref{A.4}, a
similar Chapman-Enskog expansion procedure may be considered. The zeroth and
first order LB equations Eq.\eqref{7a}, Eq.\eqref{7b} remain unchanged when
performing the Chapman--Enskog expansion, while the second-order equation %
\eqref{7c} becomes, {\setlength\arraycolsep{2pt}
\begin{eqnarray}
&&\frac{\partial }{\partial t_{2}}f_{i}^{(0)}+\frac{\partial }{\partial t_{1}%
}f_{i}^{(1)}-\frac{\delta t}{2}\frac{\partial }{\partial t_{1}}(\mathbf{S}%
_{il}f_{l}^{(1)})+v_{i\alpha }\frac{\partial }{\partial x_{1\alpha }}%
f_{i}^{(1)}+\frac{\delta t}{2}v_{i\alpha }\frac{\partial }{\partial
x_{1\alpha }}(\mathbf{S}_{il}f_{l}^{(1)}) + {}  \notag \\
&&\frac{\delta t}{2}v_{i\alpha }v_{i\beta }\frac{\partial }{\partial
x_{1\alpha }}\frac{\partial }{\partial x_{1\beta }}f_{i}^{(0)}-\frac{\delta t%
}{2}v_{i\alpha }^{2}\frac{\partial ^{2}}{\partial x_{1\alpha }^{2}}%
f_{i}^{(0)} = -\mathbf{S}_{il}f_{l}^{(2)}\text{.}  \label{A.5}
\end{eqnarray}%
} It can be converted into moment space to obtain: {\setlength%
\arraycolsep{2pt}
\begin{eqnarray}
& &\frac{\partial }{\partial t_{2}}\hat{f}_{i}^{(0)}+\frac{\partial }{%
\partial t_{1}}\hat{f}_{i}^{(1)}-\frac{\delta t}{2}\frac{\partial }{\partial
t_{1}}(\hat{\mathbf{S}}_{il}\hat{f}_{l}^{(1)})+\hat{\mathbf{E}}_{i\alpha }%
\frac{\partial }{\partial x_{1\alpha }}\hat{f}_{i}^{(1)}+\frac{\delta t}{2}%
\hat{\mathbf{E}}_{i\alpha }\frac{\partial }{\partial x_{1\alpha }}(\hat{%
\mathbf{S}}_{il}\hat{f}_{l}^{(1)}) + {}  \notag \\
& &\frac{\delta t}{2}\hat{\mathbf{E}}_{i\alpha }\hat{\mathbf{E}}_{i\beta }%
\frac{\partial }{\partial x_{1\alpha }}\frac{\partial }{\partial x_{1\beta }}%
\hat{f}_{i}^{(0)}-\frac{\delta t}{2}\hat{\mathbf{E}}_{i\alpha }^{2}\frac{%
\partial ^{2}}{\partial x_{1\alpha }^{2}}\hat{f}_{i}^{(0)} = -\hat{\mathbf{S}%
}_{il}\hat{f}_{l}^{(2)}\text{,}  \label{A.6}
\end{eqnarray}%
where $\hat{\mathbf{E}}_{\alpha }=\mathbf{M}(v_{\alpha }\mathbf{I})\mathbf{M}%
^{-1}$. From the equation, we obtain
\begin{subequations}
\begin{equation}
\frac{\partial \rho }{\partial t_{2}}+\delta t\frac{\partial }{\partial x_{1}%
}\frac{\partial }{\partial y_{1}}\hat{f}_{6}^{eq}=0\text{,}  \label{A.7a}
\end{equation}%
\begin{equation}
\frac{\partial j_{x}}{\partial t_{2}}+\frac{1}{2}\frac{\partial }{\partial
x_{1}}(1+\frac{\delta t}{2}s_{5})\hat{f}_{5}^{(1)}+\frac{\partial }{\partial
y_{1}}(1+\frac{\delta t}{2}s_{6})\hat{f}_{6}^{(1)}+\frac{\delta t}{4}\frac{%
\partial }{\partial x_{1}}\frac{\partial }{\partial y_{1}}(\hat{f}%
_{10}^{eq}+2\hat{f}_{8}^{eq})=0\text{,}  \label{A.7b}
\end{equation}%
\begin{equation}
\frac{\partial j_{y}}{\partial t_{2}}+\frac{\partial }{\partial x_{1}}(1+%
\frac{\delta t}{2}s_{6})\hat{f}_{6}^{(1)}-\frac{1}{2}\frac{\partial }{%
\partial y_{1}}(1+\frac{\delta t}{2}s_{5})\hat{f}_{5}^{(1)}+\frac{\delta t}{4%
}\frac{\partial }{\partial x_{1}}\frac{\partial }{\partial y_{1}}(2\hat{f}%
_{7}^{eq}-\hat{f}_{9}^{eq})=0\text{,}  \label{A.7c}
\end{equation}%
\begin{equation}
\frac{\partial e}{\partial t_{2}}+\frac{\partial }{\partial x_{1}}(1+\frac{%
\delta t}{2}s_{7})\hat{f}_{7}^{(1)}+\frac{\partial }{\partial y_{1}}(1+\frac{%
\delta t}{2}s_{8})\hat{f}_{8}^{(1)}+\frac{\delta t}{2}\frac{\partial }{%
\partial x_{1}}\frac{\partial }{\partial y_{1}}\hat{f}_{14}^{eq}=0\text{.}
\label{A.7d}
\end{equation}%
In this way the recovered NS equations are as follows:
\end{subequations}
\begin{subequations}
\begin{equation}
\frac{\partial \rho }{\partial t}+\frac{\partial j_{x}}{\partial x}+\frac{%
\partial j_{y}}{\partial y}=-\delta t\frac{\partial }{\partial x}\frac{%
\partial }{\partial y}(j_{x}j_{y}/\rho )\text{,}  \label{A.8a}
\end{equation}%
\begin{equation}
\frac{\partial j_{x}}{\partial t}+\frac{\partial }{\partial x}\left(
j_{x}^{2}/\rho \right) +\frac{\partial }{\partial y}\left( j_{x}j_{y}/\rho
\right) =-\frac{\partial P}{\partial x}+\frac{\partial }{\partial x}[\mu
_{s}^{^{\prime }}(\frac{\partial u_{x}}{\partial x}-\frac{\partial u_{y}}{%
\partial y})]+\frac{\partial }{\partial y}[\mu _{v}^{^{\prime }}(\frac{%
\partial u_{y}}{\partial x}+\frac{\partial u_{x}}{\partial y})]-\delta t%
\frac{\partial }{\partial x}\frac{\partial }{\partial y}[(\rho
u_{x}^{2}+P)u_{y}]\text{,}  \label{A.8b}
\end{equation}%
\begin{equation}
\frac{\partial j_{y}}{\partial t}+\frac{\partial }{\partial x}\left(
j_{x}j_{y}/\rho \right) +\frac{\partial }{\partial y}\left( j_{y}^{2}/\rho
\right) =-\frac{\partial P}{\partial y}+\frac{\partial }{\partial x}[\mu
_{v}^{^{\prime }}(\frac{\partial u_{y}}{\partial x}+\frac{\partial u_{x}}{%
\partial y})]-\frac{\partial }{\partial y}[\mu _{s}^{^{\prime }}(\frac{%
\partial u_{x}}{\partial x}-\frac{\partial u_{y}}{\partial y})]-\delta t%
\frac{\partial }{\partial x}\frac{\partial }{\partial y}[(\rho
u_{y}^{2}+P)u_{x}]\text{,}  \label{A.8c}
\end{equation}%
\begin{eqnarray}
&&\frac{\partial e}{\partial t}+\frac{\partial }{\partial x}[(e+P)j_{x}/\rho
]+\frac{\partial }{\partial y}[(e+P)j_{y}/\rho ]  \notag \\
&=&\frac{\partial }{\partial x}[\lambda _{1}^{^{\prime }}(2\frac{\partial T}{%
\partial x}+u_{y}\frac{\partial u_{y}}{\partial x}+u_{x}\frac{\partial u_{x}%
}{\partial x}-u_{x}\frac{\partial u_{y}}{\partial y}+u_{y}\frac{\partial
u_{x}}{\partial y})]  \notag \\
&&+\frac{\partial }{\partial y}[\lambda _{2}^{^{\prime }}(2\frac{\partial T}{%
\partial y}+u_{x}\frac{\partial u_{x}}{\partial y}-u_{y}\frac{\partial u_{x}%
}{\partial x}+u_{x}\frac{\partial u_{y}}{\partial x}+u_{y}\frac{\partial
u_{y}}{\partial y})]  \notag \\
&&-\delta t\frac{\partial }{\partial x}\frac{\partial }{\partial y}[(3P+%
\frac{1}{2}\rho u^{2})u_{x}u_{y}]\text{,}  \label{A.8d}
\end{eqnarray}%
where
\end{subequations}
\begin{equation*}
\mu _{s}^{^{\prime }}=\rho RT(\frac{1}{s_{5}}+\frac{\delta t}{2})\text{, }%
\mu _{v}^{^{\prime }}=\rho RT(\frac{1}{s_{6}}+\frac{\delta t}{2})\text{, }%
\lambda _{1}^{^{\prime }}=\rho R^{2}T(\frac{1}{s_{7}}+\frac{\delta t}{2})%
\text{, }\lambda _{2}^{^{\prime }}=\rho R^{2}T(\frac{1}{s_{8}}+\frac{\delta t%
}{2})\text{.}
\end{equation*}%
When $\delta t$ approaches $0$, equations \eqref{A.8a}-\eqref{A.8d}
reduce to the Eqs.\eqref{15a}-\eqref{15d}. }


\end{document}